\documentstyle[prl,aps,multicol,epsf]{revtex}

\begin{document}
\draft
\widetext
\title{Inverse photoemission in strongly correlated electron systems}
\author{R. Eder$^1$ and Y. Ohta$^2$}

\address{$^1$ Department of Applied and Solid State Physics,
University of Groningen, 9747AG Groningen, The Netherlands\\
$^2$ Department of Physics, Chiba University, Chiba 263, Japan}

\maketitle

\begin{abstract}
\leftskip 54.8pt
\rightskip 54.8pt
Based on exact results for small clusters of $t$$-$$J$ model we point
out the existence of several distinct channels in the inverse
photoemission (IPES) spectrum. Hole-like quasiparticles can either be
annihilated completely, or leave behind a variable number of spin
excitations, which formed the dressing cloud of the annihilated hole.
In the physical parameter regime the latter processes carry the bulk
of IPES weight and although the Fermi surface takes the form of hole
pockets, the distribution of spectal weight including the `spin excitation
bands' is reminiscent of free electrons.
\end{abstract}


\begin{multicols}{2}
\narrowtext
A frequently used concept in many-body physics is the `dressing' of a
particle by an excitation cloud. Sudden annihilation of the
particle then may leave behind its dressing cloud,
and these `released' excitations may carry excess momentum and energy.
Such  `shake-off' processes represent
a channel for particle annihilation of pure many-body origin, and one
may expect that their relative importance
depends on the `degree of dressing',
as measured by the quasiparticle weight $Z$. To see this, let us consider
the change of the single particle spectral function upon removing an
(infinitesimal) fraction $\epsilon$ of electrons/spin direction.
In a noninteracting system thereby the
topmost $N\epsilon$ poles of the photoemission (PES) spectrum
cross to the inverse photoemission (IPES) spectrum
($N$ denotes the system size). In a Fermi liquid the same will
happen; depending on the magnitude of $Z$, however, the corresponding
decrease of the integrated PES weight of $N\epsilon\cdot Z$ may be
substantially less than the change of $N\epsilon$ required to maintain
the correct electron count (the integrated PES weight equals the number
of electrons). Spectral weight corresponding to $N\epsilon$$(1$$-$$Z)$
electrons therefore must shift from PES to IPES
at momenta and energies off the Fermi
surface, and a natural mechanism to accomplish this
would be the `shake-off processes' adressed above.
Values of $Z$$\approx$$0.2-0.5$, as found in
the $t$$-$$J$ and Hubbard model near half-filling\cite{Dagoreview},
thus suggest that in this case in fact the bulk of IPES weight is
carried by shake-off processes.
In this manuscript we present diagonalization results which
substantiate this general argument.
The emerging picture of the single particle
spectral function and Fermi surface not only resolves the
apparent contradiction between mounting evidence\cite{rigid,pockets} for
a hole pocket-like Fermi surface in these models on one hand
and the roughly free electron-like weight distribution in
their spectral function\cite{spectra} on the other; it is also reasonably
consistent with experiments on cuprate superconductors.\\
For the standard  $t$$-$$J$ model\cite{Dagoreview}
we study the electron addition spectrum
\[
A_{+}(\bbox{k},\omega) =
\sum_\nu
| \langle \nu_{n} | \hat{c}_{\bbox{k},\uparrow}^\dagger
| 0_{n+1} \rangle |^2
\delta(\omega - \omega_\nu^{n})
\]
where $|\nu_{n} \rangle $ denotes
the $\nu^{th}$ eigenstate with
$n$ holes (in particular $\nu$$=$$0$ implies the ground state (GS))
and $\omega_\nu^{n}$ its excitation energy relative to
the $n$-hole GS.
The operator $\hat{c}_{\bbox{k},\sigma}^\dagger$ is the Fourier
transform of $\hat{c}_{i,\sigma}^\dagger$$=$$
c_{i,\sigma}^\dagger(1$$-$$n_{i,-\sigma})$.
where $c_{i,\sigma}^\dagger$ denotes the Fermion creation operator.
For small clusters $A_+(\bbox{k},\omega)$ can be evaluated exactly by the
Lanczos algorithm, the results presented below have been obtained in the
standard $18$-site cluster\cite{Dagoreview}, results for the $16$ site
cluster are consistent with these.\\
As a first step we consider the dependence of $A_{+}(\bbox{k},\omega)$
on $J/t$ and hole number $n_h$. A trivial
dependence on $n_h$ is due the requirement that the
$\bbox{k}$ and $\omega$ integrated IPES weight equals $n_h$.
Figure 1 therefore shows the IPES spectra divided by $n_h$
for the $1$ and $2$ hole ground states and various $J/t$.
The single hole GS is $8$-fold
degenerate, because it has finite momentum,
$\bbox{k}_0=(\pm2\pi/3,0), (0,\pm2\pi/3)$,
and $z$-spin $S_z$$=$$\pm1/2$; the single hole spectra of Figure 1
have been obtained by averaging over these $8$ degenerate states.
When energies are measured in units of $J$,
it is immediately obvious that apart from an overall
slight shift to lower energies in the $2$-hole case
(which may reflect the `softening' of the
spin excitations due to increased hole doping)
the two spectra are remarkably similar: they show
`features' at comparable energies, with comparable
spectral weight and with an anlogous dependence of the weight on $J/t$
(analogous features can also be
seen in the IPES spectra of the $16$-site cluster).
Up to the `softening' of peak energies, the IPES spectrum
thus scales with $n_h$ over its entire width, in strong contrast e.g. to
noninteracting electrons, where the parts far from the
Fermi energy remain unaffected by a change of electron density.
We thus see the same remarkable continuity
with hole doping as established previously
for the PES spectrum\cite{rigid} and the dynamical charge
correlation function\cite{EderOhtaMaekawa}.\\
We proceed to a detailed examination of the
single hole case. More precisely, we study the addition of an
$\uparrow$-spin electron to the single-hole GS
with momentum $(-2\pi/3,0)$ and $z$-spin $-$$1/2$;
the final states are eigenstates
of the undoped antiferromagnet, i.e. spin excited states,
with $z$-spin $S_z$$=$$0$ and total spin $S_{tot}$$=$$0,1$.
We define the spin excitation operators $S_{1,j}$$=$$S^z_j$
and $S_{2,j}$$=$$\bbox{S}_j \cdot \bbox{S}_{j+ \hat{x}}$$+$$
\bbox{S}_j \cdot \bbox{S}_{j+ \hat{y}}$
(where e.g. $j+\hat{x}$ denotes the nearest neighbor of site $j$
in $x$-direction) and their Fourier transforms $S_{\lambda,\bbox{q}}$.
Then, the following off-diagonal Green's functions
describe the interference between IPES
in the $n+1$ hole GS
and spin excitation of the $n$ hole GS:
\[
B_\lambda(\bbox{k},\omega)
= \sum_{\nu,\bbox{q}\neq 0}
\frac{
\langle 0_{n} |S_{\lambda,\bbox{q}}^\dagger | \nu_{n}\rangle
\langle \nu_{n} | c_{\bbox{k},\sigma}^\dagger|0_{n+1} \rangle}
{\langle 0_n| S_{\lambda,\bbox{q}}^\dagger
S_{\lambda,\bbox{q}}|0_n\rangle^{1/2} }
\delta(\omega - \omega_\nu^{n}).
\]
We have normalized the state $S_{\lambda,\bbox{q}}|0_n\rangle$ to unity
in order to suppress a possibly strong $\bbox{q}$
dependence of its norm;
in evaluating $B_\lambda(\bbox{k},\omega)$
we moreover readjusted for each $\bbox{k}$
the (arbitrary) relative phase between
the ground states $|0_0\rangle$ and $|0_{1}\rangle$
such that the frequency integral of
$B_\lambda(\bbox{k},\omega)$ is real and positive.
The imaginary part then vanishes identically.
Figure 2a shows $A_+(\bbox{k},\omega)$ and
$|\Re\; B_\lambda(\bbox{k},\omega)|$.
In the IPES spectrum $A_+(\bbox{k},\omega)$
there is precisely one peak with
excitation energy $\omega_\nu$$=$$0$ at the position of the `hole
momentum', $(2\pi/3,0)$. The corresponding final state is the
GS of the undoped system, so that this peak obviously describes
the inverse process of hole creation in the half-filled GS.
In a single particle picture, it would be the only peak
expected. In the actual IPEs spectrum, however, the bulk of weight is
carried by peaks with finite excitation energy and
comparison with
$B_\lambda(\bbox{k},\sigma)$ establishes a hierarchy
of increasing spin excitation:
the lowest group of the finite energy peaks corresponds to
a single spin wave left behind in the cluster, the
next highest group to a `bimagnon' and
most probably even higher peaks correspond to an even stronger
spin excitation left behind by the annihilated hole.
As regards the Fermi surface, it is clear that
only the lowest state in this hierarchy,
corresponding to `complete annihilation'
of the quasiparticle should be considered. The IPES spectrum
thus consists of the
`hole pocket' at $(2\pi/3,0)$\cite{pockets}
plus the `shake off bands' at higher energies.
The latter are present
throughout the Brillouin zone, however with a strongly
reduced spectral weight near $\Gamma$; this is to be expected
from the kinetic energy sum rule
$E_{kin}=-\sum_{\bbox{k}} \epsilon_{\bbox{k}} W^{IPES}_{\bbox{k}}$,
where $\epsilon_{\bbox{k}}$ denotes the noninteracting kinetic energy
and $W^{IPES}_{\bbox{k}}$ the integrated IPES weight for momentum
$\bbox{k}$: large (small) IPES weight near the zone boundary (zone center)
optimizes the kinetic energy.\\
We proceed to the two-hole case. The GS in this case
is a spin singlet and has $d_{x^2-y^2}$ symmetry;
introducing $\bbox{k}_1=(2\pi/3,0)$ and $\bbox{k}_2=(0,2\pi/3)$,
the simplest two-particle state with these quantum numbers
to be expected within a rigid-band picture would be:
\begin{eqnarray}
|\Psi_0\rangle &=&
\frac{1}{2} [\;
a_{k_1,\uparrow}^\dagger
a_{-k_1,\downarrow}^\dagger -
a_{k_1,\downarrow}^\dagger
a_{-k_1,\uparrow}^\dagger
\nonumber \\
&\;&\;\;\;\;\;\;-
a_{k_2,\uparrow}^\dagger
a_{-k_2,\downarrow}^\dagger +
a_{k_2,\downarrow}^\dagger
a_{-k_2,\uparrow}^\dagger
\;] |vac \rangle.
\end{eqnarray}
For this state, the `quasiparticle occupation'
of the $4$ equivalent momenta $\pm \bbox{k}_1$, $\pm\bbox{k}_2$
is the same for each spin
direction. Thus if we make the simplest assumption possible,
namely that the annihilation of one
hole is not at all influenced by the presence of
the second one, the two-hole IPES spectrum should simply
be the average of the single hole IPES spectra for the different
momenta and spin directions, as is indeed the case.
Figure 2b shows the `interference spectrum'
$|\Re B_\lambda(\bbox{k},\omega)|$ for $n$$=$$1$
(the latter spectrum being averaged over the $4$ degenerate
single hole ground states with spin $1/2$).
Unlike the $n$$=$$0$ case, the IPES peaks now
show overlap with both types of spin excitations - this is to be expected,
because the extra spin defects generated by the
operators $S_{\lambda,\bbox{q}}$ in the single hole GS may also be
`absorbed' into the dressing cloud of the hole.
For example, addition of a spin excitation with momentum $(2\pi/3,2\pi/3)$
to the single hole GS with momentum $(-2\pi/3,0)$
with a certain probability
may simply give the degenerate GS with momentum $(0,2\pi/3)$.
Apart from this slight complication, however, the structure of
$B_\lambda(\bbox{k},\omega)$ is quite similar to the single hole case:
the single spin wave operator $S_1$ has strongest overlap with the dominant
peak at the bottom of the respective IPES spectrum,
whereas the smaller IPES peaks at higher energies rather
correspond to the `bimagnon': obviously there is the same
hierarchy of increasing spin excitation as in the single hole case
(as is to be expected from the simple scaling of the IPES spectrum
with $n_h$, compare Figure 1.\\
Taken together, the numerical data suggest the
existence of different `channels' in the IPES spectrum:
in addition to a `conventional channel' at the lowest excitation energies,
where a hole-like quasiparticle is `annihilated completely',
there exist higher energy final states where the annihilation
process of the quasiparticle
leaves behind a variable number of spin excitations, which
formed the dressing cloud of the annihilated hole. The high degree of
continuity of the IPES spectrum with doping moreover suggests that the
spin-bag like quasiparticles for the half-filled case
persist as well-defined entities
also in the two-hole case (i.e. at nominal hole concentrations
$\delta$$\approx$$12$\%).
Assuming that these different channels remain `disconnected'
in the infinite system, i.e. that they do not merge to form
a single `band', the Fermi surface therefore should take the form
of hole pockets, consistent with increasing numerical
evidence\cite{pockets}. The emerging picture of the
low energy weight distribution and Fermiology in the full single particle
spectral function then is summarized in Figure 3a
(thereby the incoherent continua deep below
the Fermi energy are omitted).
To begin with, the Fermi surface is a hole pocket\cite{pockets}
generated by shifting the chemical potential into the
more or less rigid
next-nearest neighbor hopping band for hole motion in the undoped
antiferromagnet\cite{rigid}.
Thereby the PES spectral weight of the quasiparticle band in the outer
parts of the Brillouin zone is
small\cite{rigid,Dagoshadow}; this `shadow band'\cite{KampfSchrieffer}
effect is readily understood\cite{rigid}
by the corollary of the kinetic energy sum-rule,
$E_{kin}$$=$$\sum_{\bbox{k}} \epsilon_{\bbox{k}} W^{PES}_{\bbox{k}}$,
with $W^{PES}_{\bbox{k}}$ the integrated PES weight.
Low weight in the shadow band allows on one hand to
maintain the next-nearest neighbor dispersion of width $J$, which
seems to be optimal for the exchange mediated
propagation of a single hole\cite{Dagoreview},
while at the same time staying close to the free-electron
$W^{PES}_{\bbox{k}}$ (see Figure 3a)
so as to optimize the `ordinary' kinetic energy.
We turn to the IPES spectrum, which consists of several components:
the hole pockets right at $E_F$
near $(\pi/2,\pi/2)$ (or $(\pi,0)$ for two holes in
a small clusters\cite{rigid}) and the various `shake-off bands'
at higher energy and predominantly in the outer part of the Brillouin zone.
When viewed with coarse energy and/or $\bbox{k}$
resolution, the resulting distribution of spectral weight may be
rather similar to a `renormalized free electron band'\cite{spectra};
let us note that this is largely necessitated by elementary
rules\cite{comment}.
Key features which would allow for a distinction are
a) the low intensity `shadow bands' in the outer parts of the Brillouin
zone,
b) a `disconnected' IPES spectrum consisting of low energy hole pockets
near $(\pi/2,\pi/2)$ and higher energy magnon bands near
$(\pi,\pi)$, and
c) a Fermi surface with a volume proportional to the
hole concentration $\delta$.\\
To address these issues, we proceed to a comparison with experiments
on cuprate superconductors.
PES experiments on the insulating antiferromagnet
$Sr_2CuO_2Cl_2$\cite{Wells}
have given evidence for a dramatic shadow band effect
in this undoped antiferromagnet. Assuming that the weight of the
shadow bands even decreases in the doped case (as is suggested by cluster
studies\cite{rigid,Dagoshadow}) would readily explain the
non-observation of the shadow bands in early ARPES experiments.
A recent Fermi surface mapping
of the metallic compound  $Bi_2 Sr_2 Ca Cu_2 O_{8+x}$
with an extremely high density of $\bbox{k}$-points
by Aebi {\em et al.}\cite{Aebi}  has
indeed shown indications for the shadow bands also in the doped case,
a result which has been corroborated by
conventional ARPES in the meantime\cite{Ding}.\\
We turn to IPES, where experimental data are more sparse
than in the PES case.
IPES experiments which revealed several dispersive bands were
reported by Bernhoff {\em et al.}\cite{Bernhoff}.
Figure 3b shows the positions of IPES peaks found by these authors
in $Bi_2 Sr_2 Ca Cu_2 O_8$, as well as an assignment of bands
compatible with our scenario.
Grouping the higher lying peaks as indicated in the figure, one can obtain
almost quantitative agreement with
the LDA band structure\cite{Massidda}, {\em provided} the two $CuO_2$-plane
derived bands are omitted from the latter.
We are then left with
only two `disconnected' low energy features: the extreme low energy
intensity near $(\pi/2,\pi/2)$, which we
associate with the hole pocket plus some `isolated' spectral weight near
$(\pi,\pi)$. The latter feature is most clearly visible at the
largest possible distance from $\bbox{k}_F$ and seems to
disappear when approaching $\bbox{k}_F$. This is in
rather clear contrast to the behaviour expected for a
`renormalized free-electron band', which on the contrary should be
most clearly visible near $\bbox{k}_F$.
On the other hand,  this is precisely the behaviour expected for
the magnon bands in the $t$$-$$J$ model, whose large intensity near
$(\pi,\pi)$ is favourable to lower the kinetic energy.\\
We turn to the volume of the Fermi surface. Here the most direct probe
the deHaas--vanAlphen effect, which
is insensitive to the quasiparticle weight $Z$ and hence may detect
the `shadow' sheets of the Fermi surface.
Modelling the holes as spin $1/2$ particles distributed evenly over
$4$ symmetric pockets of equal size,
the fraction of the Brillouin zone covered by a single pocket is
$\delta/8$.  With a lattice constant of $3.85\AA$
(as would be appropriate for most cuprate superconductors)
$\delta$$=$$0.18$
gives a de Haas--van Alphen cross-section of $0.63 kT$,
which is roughly half-way in between two orbits of
$(0.53\pm0.02)kT$ and  $(0.78\pm0.02)kT$ actually observed in
$YBa_2Cu_3O_{6.97}$\cite{Fowler}. This suggests
to assume that
the two `$t$$-$$J$ bands' derived from the two $CuO_2$ planes/unit cell
form bonding and antibonding combinations so that there is
a disproportionation of holes between these two bands.
With  $\delta_1$$=$$0.15$, $\delta_2$$=$$0.22$ (i.e. an average
$\delta$ of $0.185$) one obtains orbits of
$0.52kT$ and $0.76kT$, consistent with experiment\cite{Fowler}.
For completeness we note that the
experimental doping dependence of the low temperature Hall constant,
which is another $Z$-independent quantity directly related to the carrier
concentration, can also be reproduced well within the
hole-pocket picture\cite{Trugman,Dagottoflat}\\.
In summary, we have presented a cluster diagonalization study of the
inverse photoemission spectrum for the low doping regime of the $t$$-$$J$
model. The results first of all show a remarkable continuity
with doping which, together with similar results for the
photoemission spectra\cite{rigid} and
dynamical correlation functions\cite{EderOhtaMaekawa}
suggests that the single hole is the key problem for understanding
the moderately doped region.
The emerging picture of the single particle
spectral function is that of a spectral weight distribution which
in the neighborhood of the Fermi surface roughly resembles
that of free electrons, combined with a hole-pocket Fermi surface
with a volume proportional to the number of doped holes.
Various experimental data
which suggest that this in cuprate superconductors
are consistent with this scenario.\\
Financial support of R. E. through the European Community is
most gratefully acknowledged. Computations were carried out at the
Institute for Molecular Science, Okazaki.
\figure{Figure 1:
        IPES spectrum $A_+(\bbox{k},\omega)/n_h$ for different $\bbox{k}$
        and $J/t$ for $n_h$$=$$1$ (a) and $n_h$$=$$2$ (b).
        $\delta$-functions are replaced by Lorentzians of width
        $0.1J$.
\label{fig1}}
\figure{Figure 2: (a) IPES spectrum $A_+(\bbox{k},\omega)$
        (full line), ``interference spectra''
        $0.5\cdot B_1(\bbox{k},\omega)$ (dashed line)
        and $B_2(\bbox{k},\omega)$ (dotted line), for $n$$=$$1$.
        The ratio $J/t$$=$$0.4$,
        $\delta$-functions are replaced by Lorentzians of width
        $0.01J$.
        (b) Same as (a) for $n$$=$$2$.
\label{fig2}}
\figure{Figure 3: (a) Schematic spectral function
         of the $t$$-$$J$ model near the Fermi energy.
         The full line denotes the PES intensity from the
         quasiparticle band, the dashed line the IPES spectrum from the
         `hole pocket', the short-dashed line the `shake-off bands'.
        (b) Comparison of experimental and theoretical band structure of
         $Bi_2 Sr_2 Ca Cu_2 O_8$.
         The Fermi level is the zero of energy, the symbols denote
         prominent peaks in the IPES spectrum\cite{Bernhoff} (above $E_F$)
         and in the PES spectrum\cite{Ding} (below $E_F$).
         Dashed lines show the unoccupied part of the
         LDA bandstructure\cite{Massidda},
         with the $CuO_2$-plane derived bands removed, and
         rescaled and shifted according to $E$$=$$0.77\cdot
         (E_{LDA}$$-$$E_F)$$-$$0.32eV$.
         The full line indicates the `$t$$-$$J$ band'. A group of PES
         peaks originating from the superlattice structure of
         $Bi_2 Sr_2 Ca Cu_2 O_8$\cite{Ding} is omitted.
\label{fig3}}

\end{multicols}
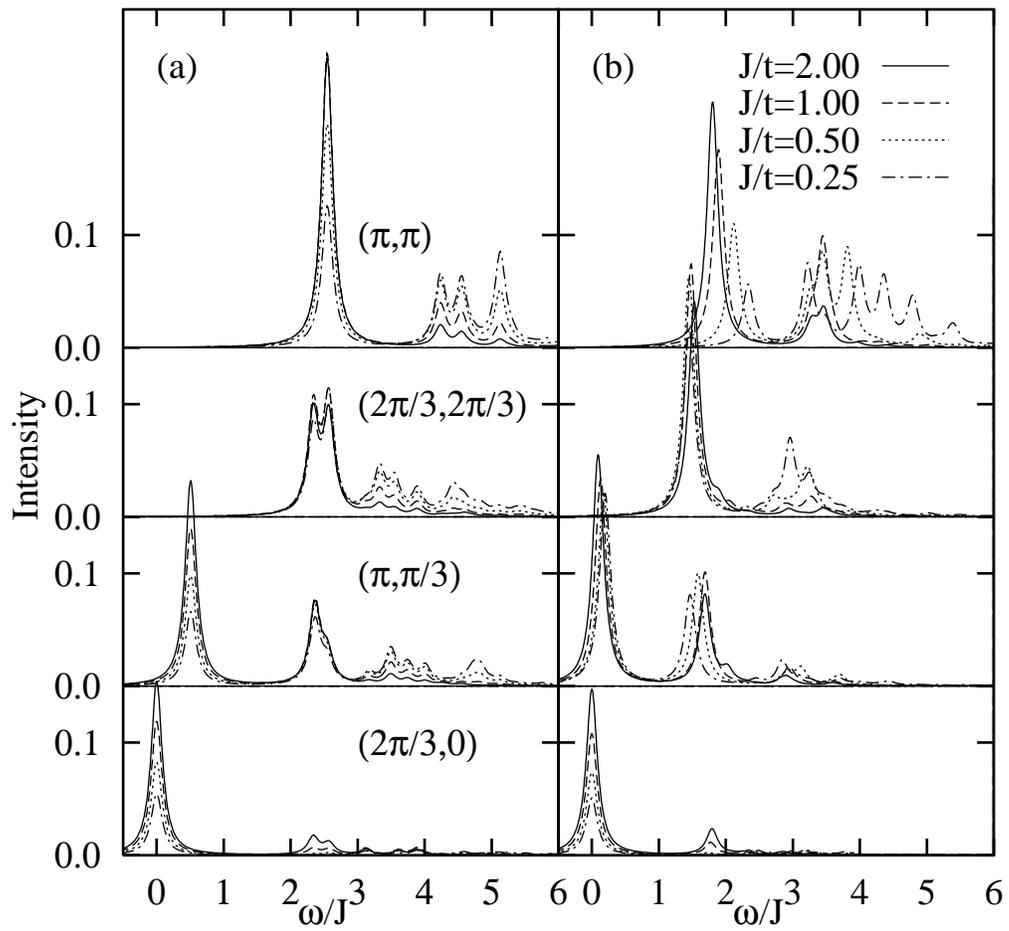
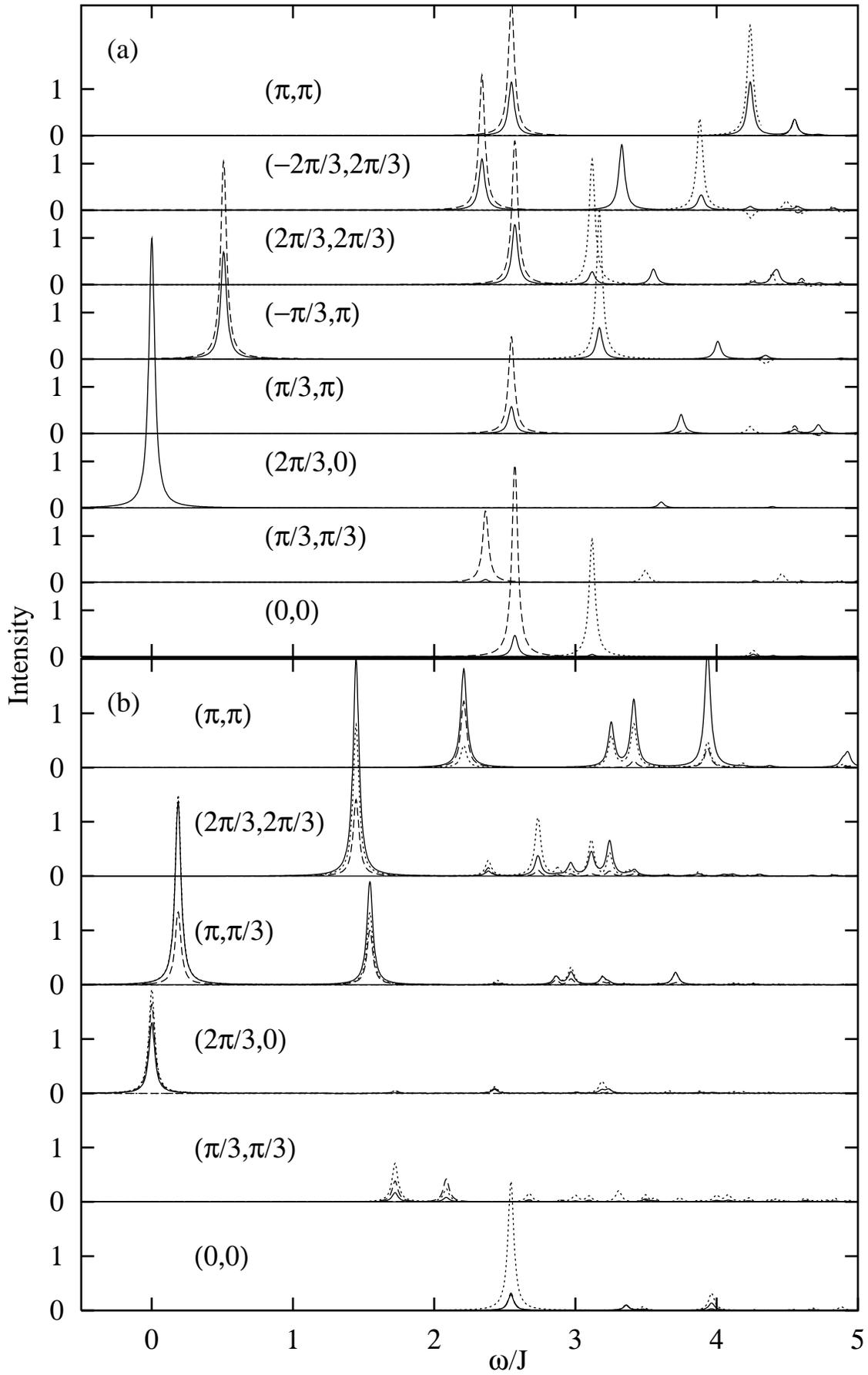
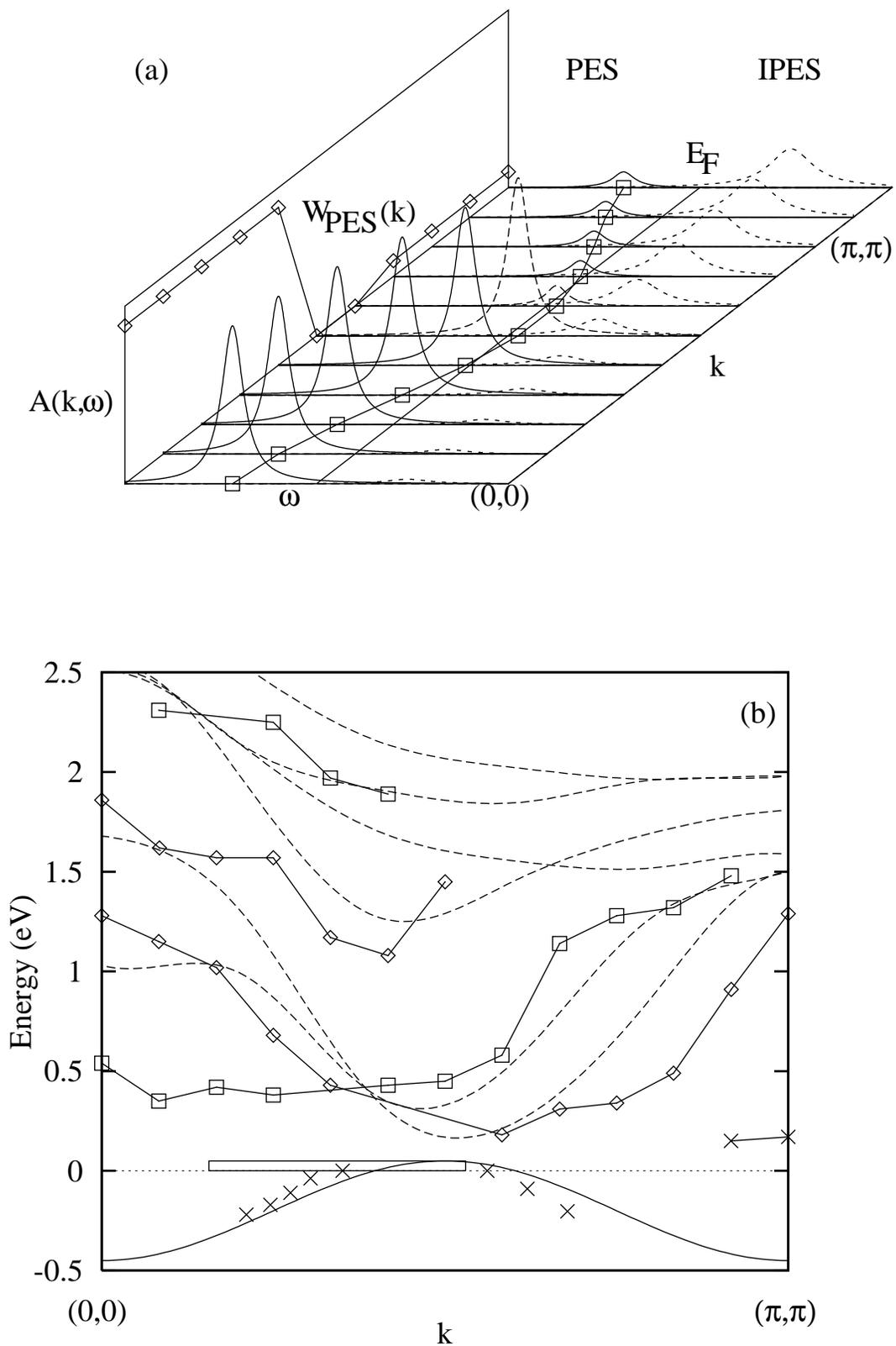
\end{document}